\documentclass[a4paper, UKenglish, cleveref, autoref, thm-restate]{lipics-v2019} \usepackage[lipics]{optional}

\usepackage[textsize=scriptsize]{todonotes}

\usepackage{sidecap}

\usepackage{gensymb}

\newcommand{\cadnano}{cadnano}

\usepackage{inconsolata}

\usepackage{xcolor}
\usepackage{cite}

\colorlet{punct}{red!60!black}
\definecolor{background}{HTML}{EEEEEE}
\definecolor{delim}{RGB}{20,105,176}
\colorlet{numb}{magenta!60!black}

\lstdefinelanguage{json}{
    basicstyle=\scriptsize\ttfamily,
    showstringspaces=false,
    breaklines=true,
    frame=lines,
    backgroundcolor=\color{background},
    literate=
     *{0}{{{\color{numb}0}}}{1}
      {1}{{{\color{numb}1}}}{1}
      {2}{{{\color{numb}2}}}{1}
      {3}{{{\color{numb}3}}}{1}
      {4}{{{\color{numb}4}}}{1}
      {5}{{{\color{numb}5}}}{1}
      {6}{{{\color{numb}6}}}{1}
      {7}{{{\color{numb}7}}}{1}
      {8}{{{\color{numb}8}}}{1}
      {9}{{{\color{numb}9}}}{1}
      {:}{{{\color{punct}{:}}}}{1}
      {,}{{{\color{punct}{,}}}}{1}
      {\{}{{{\color{delim}{\{}}}}{1}
      {\}}{{{\color{delim}{\}}}}}{1}
      {[}{{{\color{delim}{[}}}}{1}
      {]}{{{\color{delim}{]}}}}{1},
}

\lstnewenvironment{jsonlisting}
    {\lstset{language=json}}
    {}

\definecolor{codegreen}{rgb}{0,0.6,0}
\definecolor{codegray}{rgb}{0.5,0.5,0.5}
\definecolor{codepurple}{rgb}{0.58,0,0.82}
\definecolor{backcolour}{rgb}{0.95,0.95,0.92}

\lstnewenvironment{pythonlisting}
    {\lstset{
    language=Python,
    basicstyle=\scriptsize\ttfamily,
    keepspaces=true,
    backgroundcolor=\color{backcolour},
    commentstyle=\color{codegreen},
    breaklines=True,
    otherkeywords={self,True,False,yield},
    keywordstyle=\ttfamily\color{blue!90!black},
    keywords=[3]{ttk},
    keywordstyle={[2]\ttfamily\color{orange!80!orange}},
    keywordstyle={[3]\ttfamily\color{red!80!orange}},
    emph={MyClass,__init__},
    emphstyle=\ttfamily\color{red!80!black},
    stringstyle=\color{green!80!black},
    escapeinside={\#*}{*)},
    showstringspaces=false
    }}
    {}

\newcommand{\websites}{
    \noindent
    \url{https://scadnano.org},\ \ 
    \url{https://scadnano.org/dev}
    (stable/dev versions)
    \\
    \url{https://github.com/UC-Davis-molecular-computing/scadnano}
    (web interface code repository)
    \\
    \url{https://github.com/UC-Davis-molecular-computing/scadnano-python-package}
    (Python scripting library code repository)
    \\
    \url{https://scadnano-python-package.readthedocs.io}
    (Python scripting library API)
    \\
    \url{https://github.com/UC-Davis-molecular-computing/scadnano/blob/master/tutorial/tutorial.md}
    (web interface tutorial)
    \\
    \url{https://github.com/UC-Davis-molecular-computing/scadnano-python-package/blob/master/tutorial/tutorial.md}
    (Python scripting library tutorial)
}

\opt{lipics} {\title{scadnano: A browser-based, scriptable tool for designing DNA nanostructures} 

\titlerunning{scadnano: A browser-based, scriptable tool for designing DNA nanostructures} 

\author
{David Doty\footnote{Corresponding author}}
{University of California, Davis, USA \and \url{https://web.cs.ucdavis.edu/~doty/}}
{doty@ucdavis.edu}
{https://orcid.org/0000-0002-3922-172X}
{Supported by NSF grants 1619343, 1900931, and CAREER grant 1844976.}

\author
{Benjamin L Lee}
{University of California, Davis, USA}
{bnllee@ucdavis.edu}
{https://orcid.org/0000-0003-2307-075X}
{Supported by REU supplement through NSF CAREER grant 1844976.}

\author
{Tristan St\'{e}rin}
{Maynooth University \and \url{https://dna.hamilton.ie/tsterin/index.html}}
{Tristan.Sterin@mu.ie}
{https://orcid.org/0000-0002-2649-3718}
{Supported by European Research Council (ERC) under the European Union’s Horizon 2020 research and innovation programme (grant agreement No 772766, Active-DNA project), and Science Foundation Ireland (SFI) under Grant number 18/ERCS/5746.}

\authorrunning{D. Doty, B. Lee, T. St\'{e}rin} 

\Copyright{David Doty, Benjamin L Lee, Tristan St\'{e}rin} 



\category{} 

\relatedversion{} 

\supplement{
    \websites
} 

\acknowledgements{We thank Matthew Patitz for beta-testing and feedback, and Pierre-\'{E}tienne Meunier, author of codenano, for valuable discussions regarding the data model/file format. We are grateful to anonymous reviewers whose detailed feedback has increased the presentation quality.}

\nolinenumbers 

\hideLIPIcs  

\EventEditors{Matthew J. Patitz and Cody Geary}
\EventNoEds{2}
\EventLongTitle{26th Conference on DNA Computing and Molecular Programming (DNA 2020)}
\EventShortTitle{DNA 2020}
\EventAcronym{DNA}
\EventYear{2020}
\EventDate{September 13--18, 2020}
\EventLocation{Oxford, United Kingdom}
\EventLogo{}
\SeriesVolume{}
\ArticleNo{}
}
\opt{article}{\input{author_title_headers_article}}

\begin{document}

\maketitle

\begin{abstract}
    We introduce \emph{scadnano}
    (short for ``scriptable cadnano''),
    a computational tool for designing synthetic DNA structures.
    Its design is based heavily on \cadnano~\cite{douglas2009rapid},
    the most widely-used software for designing DNA origami~\cite{rothemund2006folding},
    with three main differences:

    \begin{enumerate}
        \item
        scadnano runs entirely in the browser, with \emph{no software installation} required.

        \item
        scadnano designs, while they can be edited manually, can also be created and edited by a \emph{well-documented Python scripting library},
        to help automate tedious tasks.

        \item
        The scadnano file format is \emph{easily human-readable}.
        This goal is closely aligned with the scripting library,
        intended to be helpful when debugging scripts or interfacing with other software.
        The format is also somewhat more \emph{expressive} than that of \cadnano,
        able to describe a broader range of DNA structures than just DNA origami.
    \end{enumerate}
\end{abstract}

\clearpage
\pagenumbering{arabic}

\section{Introduction}
\label{sec:intro}

\subsection{DNA origami and \cadnano}
Since its inception almost 15 years ago,
DNA origami~\cite{rothemund2006folding} has stood as the most reliable, high-yield, and low-cost method for synthesizing uniquely addressed DNA nanostructures,
on the order of $100$~nm wide, with $\approx 6$~nm addressing resolution
(i.e., that's how far apart individual strands are).\footnote{
    The basic idea of DNA origami is to use a long \emph{scaffold} strand (either synthesized or natural; the most common choice is the natural circular single-stranded virus known as M13mp18, 7249 bases long),
    and to synthesize shorter (a few dozen bases long) \emph{staple} strands designed to bind to multiple regions of the scaffold.
    Upon mixing in standard DNA self-assembly buffer conditions (e.g., 10 mM Tris, 1 mM EDTA, pH 8.0, 12.5~mM MgCl$_2$),
    with staples ``significantly'' more concentrated than the scaffold
    (typical concentrations are 1~nM scaffold and 10~nM each staple),
    and annealing from 90\degree C to 20\degree C for one hour,
    the staples bind to the scaffold and fold it into the desired shape, while excess staples remain free in solution and are easily separated from the formed structures by standard purification techniques.
}
To create the original designs, Rothemund wrote custom Matlab scripts to generate and visualize the designs (with ASCII art).
Soon after, the software \cadnano\ was developed by Douglas et al.~\cite{douglas2009rapid}, as part of a project extending the original 2D DNA origami results to 3D structures~\cite{douglas2009self}.
\cadnano\ has become a standard tool in structural DNA nanotechnology,
used for describing most major DNA origami designs.

\subsection{scadnano}

\begin{figure}[h]
\begin{center}
\includegraphics[width=0.95\textwidth]{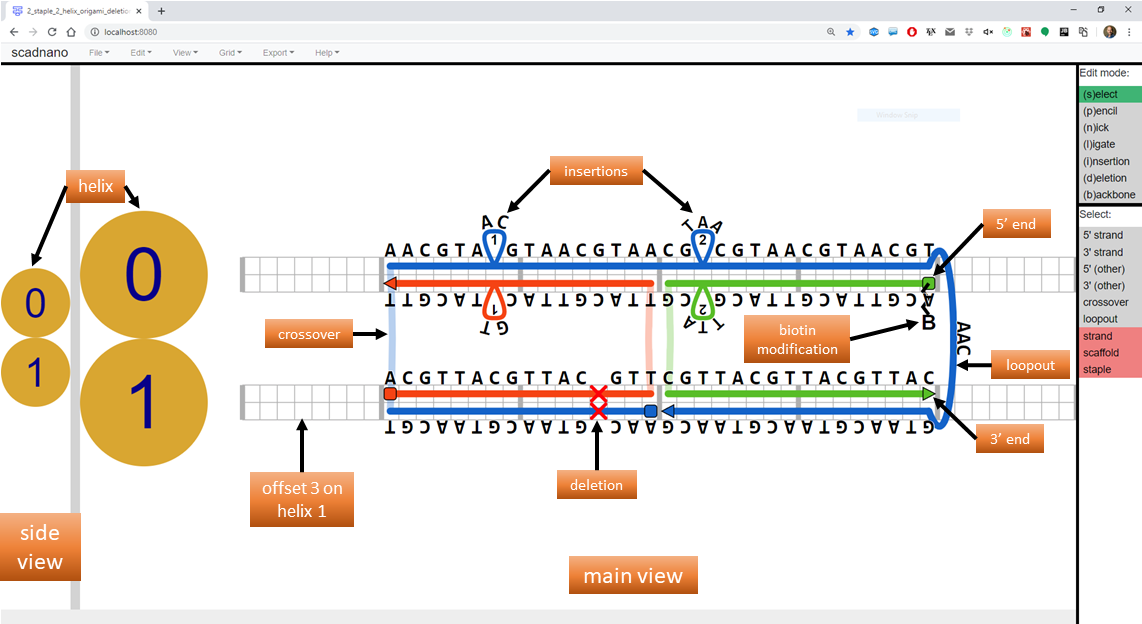}
\caption{
    screenshot of scadnano,
    annotated with some labels (in orange rectangles) to point out various parts of the data model.\footnotemark\
    The center part is the \emph{main view}, which shows the $x$ and $y$ coordinates;
    most editing takes place here.
    On the left is the \emph{side view}, which shows the $z$ and $y$ coordinates.
    $y$ increases going \emph{down} in both views (so-called ``screen coordinates),
    $x$ increases going \emph{right} in the main view and going \emph{into} the screen in the side view.
    $z$ increases going \emph{right} in the side view and going \emph{out of} the screen in the main view.
    The Edit modes on the right change what sorts of edits are possible, and the Select modes change what sort of objects can be selected while in the ``select'' edit mode.
}
\label{fig:screenshot-initial}
\end{center}
\end{figure}

The scadnano graphical interface is shown in \cref{fig:screenshot-initial}; it mimics that of \cadnano.

The goal of scadnano is to aid in designing large-scale DNA nanostructures, such as DNA origami, with ability to edit structures either manually, or programmatically through a scripting library.
scadnano seeks to imitate most of the features of \cadnano,
with three major differences that enhance the usability and interoperability of scadnano:

\begin{enumerate}
    \item
    scadnano runs entirely in the browser, with \emph{no software installation} required.
    It aims, above all else, to be simple and easy to use, well-suited for teaching, for example.

    \item
    scadnano designs, while they can be edited manually, can also be created and edited by a \emph{well-documented Python scripting library},
    to help automate tedious tasks.\footnote{
        \cadnano\ v2.5 has a Python scripting library, but its documentation is incomplete~\cite{cadnano_two_five_api}, and \cadnano\ v2.5 has not been updated for two years~\cite{cadnano_two_five} at the time of this writing.
    }

    \item
    The scadnano file format is \emph{easily human-readable} and expressive,
    natural for describing a broader range of DNA structures than just DNA origami.
    This goal is closely aligned with the scripting library,
    useful when debugging scripts or interfacing with other software.
    A related project, codenano~\cite{codenano}, uses essentially the same file format,
    developed simultaneously in consultation with the main author of codenano.
\end{enumerate}

The major features of scadnano are described in more detail in \cref{sec:features}.
Designed with interoperability in mind, any cadnano design can be imported into scadnano,
and scadnano designs obeying certain constraints
(see \cref{sec:file_formats})
can be exported to cadnano.

\subsection{Related work}

\cadnano\ is the most related prior work, and its design was the inspiration for scadnano.
\cref{subsec:features:shared_with_cadnano} goes into detail about features that scadnano shares in common with \cadnano, and the rest of \cref{sec:features} discusses some extra features in scadnano.
codenano is close in purpose to scadnano~\cite{codenano},
being also browser-based and scriptable.
Unlike scadnano, codenano includes 3D visualisation components
but not graphical editing.

vHelix~\cite{vhelix} offers comprehensive 3D origami editing and visualisation features but relies on Autodesk Maya.
Adenita~\cite{adenita} is a design and visualisation tool that allows one to work with various DNA nanostructures:
standard parallel-helix DNA origami, wireframe origamis \cite{athena}, and tile-based designs.
Adenita is distributed within the SAMSON~\cite{samson} molecular modeling platform.
Specific to the domain of 2D and 3D wireframe origamis, ATHENA \cite{athena} provides both an editing interface and sequence design algorithms that generate staple sequences from a 2D sketch.
Not related to graphical or script-based DNA design editing, the following software provides structural prediction tools for various features of DNA designs:
CanDo \cite{cando} (finite elements-based 3D structure prediction),
NUPACK and ViennaRNA \cite{nupack,viennaRNA} (thermodynamic energy of DNA strands),
oxDNA \cite{oxDNA} (kinetics prediction by molecular dynamics simulation),
and
MrDNA \cite{mrdna} (3D structure and kinetics prediction).

\footnotetext{This design is intended merely to show some scadnano features, not to show proper design respecting DNA crossover geometry; it would be strained if actually assembled.}

\subsection{Paper outline}
\cref{sec:data_model_and_file_format} describes the data model used by scadnano to represent a DNA design, and its closely related storage file format,
including a comparison with \cadnano's file format.
\cref{sec:features} describes several features of scadnano, including some that are absent from \cadnano.
\cref{sec:architecture} explains the software architecture of scadnano.
\cref{sec:architecture} is not necessary to understand how to use scadnano,
but it helps to justify why scadnano may be simpler to maintain and enhance in the future.
\cref{sec:conclusion} discusses possible future features.

This paper is not a self-contained document describing scadnano in full.
See the supplementary material links for online documentation, tutorials, and the Python library API.

\section{Data model and file format}
\label{sec:data_model_and_file_format}

\newcommand{\field}[1]{\lstinline{#1}}

\subsection{scadnano data model}
\label{subsec:scadnano_data_model_and_file_format}


Although scadnano and its data model are natural for describing DNA origami, it can be used to describe any DNA nanostructure composed of several DNA strands.
Like \cadnano,
scadnano is especially well-suited to structures where all DNA helices are parallel,
which includes not only origami, 
but also certain
tile-based designs (e.g.,~\cite{winfree1998design, SST2D, drmaurdsa}),
or ``criss-cross slat'' assembly~\cite{crisscross_assembly}.
The basic concepts, explained in more detail below,
are that the design is composed of several \emph{strands}, 
which are bound to each other on some domains, 
and possibly single-stranded on others, 
and double-stranded portions of DNA occupy a \emph{helix}.

\paragraph*{DNA Design}
An example DNA design is shown in \cref{fig:screenshot-initial}, showing most of the features discussed here.
A design (the type of object stored in a {\tt .sc} file produced when clicking ``Save'' in scadnano) consists of a \field{grid} type (a.k.a., \emph{lattice}, one of the following types: \field{square}, \field{honeycomb}, \field{hex}, or \field{none}, explained below), a list of \field{helices}, and a list of \field{strands}. 
The order of strands in the list generally doesn't matter, although it influences which are drawn on top, so a strand later in the list will have its crossovers drawn over the top of earlier strands.

\paragraph*{Helices}
Unlike strands,
the order of the helices matters; 
if there are $h$ helices, 
the helices are numbered 0 through $h-1$.
This can be overridden by specifying a field called {\tt idx} in each helix, 
but the default is to number them consecutively.
Each helix defines a set of integer \emph{offsets} with a minimum and maximum; in the example above, the minimum and maximum for each helix are 0 and 48, respectively, so 48 total offsets are shown.
Each offset is a position where a DNA base of a strand can go.

Helices in a grid (meaning one of square, honeycomb, or hex) have a 2D integer \field{grid\_position} depicted in the side view (see \cref{fig:Grids}).
Helices without a grid (meaning grid type none) have a \field{position}, a 3D real vector describing their \field{x}, \field{y}, \field{z} coordinates.
Each Helix also has fields to describe angular orientation,
using the ``aircraft principle axes'' \field{pitch}, \field{roll}, and \field{yaw} (default 0), although this feature is currently 
not well-supported 
(\url{https://github.com/UC-Davis-molecular-computing/scadnano/issues/39}).
The coordinates of helices in the main view depends on \field{grid\_position} if a grid is used, and on \field{position} otherwise. 
(Each grid position is essentially interpreted as a position with \field{z} = \field{pitch} = \field{roll} = \field{yaw} = 0.)
Helices are listed from top to bottom in the order they appear in the sequence, 
unless the property \field{helices\_view\_order} is specified in the design to display them in a different order, though currently this can only be done in the scripting library.

Helix.\field{roll} describes the DNA backbone rotation about the long axis of the helix.
At the offset \field{Helix.min\_offset}, the backbone of the forward strand on that helix has angle \field{Helix.roll}, 
where we define 0 degrees to point to straight \emph{up} in the side view. 
Rotation is \emph{clockwise} as the rotation increases from 0 up to 360 degrees.
This feature is  not intended as a globally predictive model of stability.
Rather, it helps visualize backbone angles, 
to place crossovers that minimize strain, 
by ensuring crossovers are ``locally consistent'',
without enforcing a global notion of absolute backbone rotation on all offsets in the system.

\paragraph*{Strands and domains}
Each strand is defined primarily by an ordered list of \field{domains}.
Each domain is either a single-stranded \emph{loopout} not associated to any helix, or it is a \emph{bound domain}: a region of the strand that is contiguous on a single helix.
The phrase is a bit misleading, since a bound domain is not necessarily bound to another strand, but the intention is for most of them to be bound, and for single-stranded regions usually to be represented by loopouts.

Each bound domain is specified by four mandatory properties:
\field{helix} (indicating the index of the helix on which the domain resides),
\field{forward}
(a direction can be forward or reverse, indicated by whether this field is true or false),
\field{start} integer offset, 
and a larger \field{end} integer offset.
As with common string/list indexing in programming languages, \field{start} is inclusive but \field{end} is exclusive.
So for example, a bound domain with \field{end}=8 is adjacent to one with \field{start}=8.
In the main view, \field{forward} bound domains are depicted on the top half of the helix, and \emph{reverse} (those with \field{forward}=false) are on the bottom half.
If a bound domain is forward, then \field{start} is the offset of its 5' end, and \field{end}$-1$ is the offset of its 3' end, 
otherwise these roles are reversed.
There is implicitly a crossover between adjacent bound domains in a strand.
Loopouts are explicitly specified as a (non-bound) domain in between two bound domains.
Currently, two loopouts cannot be consecutive (and this will remain a requirement),
and a loopout cannot be the first or last domain of a strand (this may be relaxed in the future).

Bound domains may have optional fields, notably \field{deletions} (called \emph{skips} in cadnano) and \field{insertions} (called \emph{loops} in cadnano).
They are a visual trick used to allow bound domains to appear to be one length in the main view of scadnano, while actually having a different length. 
Normally, each offset represents a single base. 
If instead a deletion appears at that offset, then it does not correspond to any DNA base. 
If an insertion appears at that offset, it has a positive integer \field{length}: the number of bases represented by that offset is \field{length}$+1$. 

\paragraph*{Strand optional fields}

Each strand also has a \field{color} and a Boolean field \field{is\_scaffold}.
DNA origami designs have at least one strand that is a scaffold (but can have more), and a non-DNA-origami design is simply one in which every strand has \field{is\_scaffold} = false.
Unlike cadnano, a scaffold strand can have either direction on any helix.
When there is at least one scaffold, all non-scaffold strands are called \emph{staples}.
The general idea behind DNA origami is that 
all binding is between scaffolds and staples, 
never scaffold-scaffold or staple-staple.
However, this convention is not enforced by scadnano; there are legitimate reasons for non-scaffold strands to bind to each other (e.g., DNA walkers~\cite{gu2010proximity} or circuits~\cite{chatterjee2017spatially} on the surface of an origami).

A strand can have an optional DNA \field{sequence}.
Of course, since the whole point of this software is to help design DNA structures, at some point a DNA sequence should be assigned to some of the strands.
However, it is often best to mostly finalize the design before assigning a DNA sequence, which is why the field is optional.
Many of the operations attempt to keep things consistent when modifying a design where some strands already have DNA sequences assigned, but in some cases it's not clear what to do. 
(e.g., what DNA sequence results when a length-5 strand with sequence AACGT is extended to be longer?)

\paragraph*{DNA modifications}
DNA modifications describe ways that various small molecules may be attached to synthetic DNA as part of the DNA synthesis process.
Common DNA modifications include biotin 
(useful for binding to the protein streptavidin)
and fluorophores such as Cy3 
(useful for light microscopy).
Modifications can be attached to the 5' end, the 3' end, or to an internal base.

A few pre-defined modifications are provided as examples 
in the Python scripting library.
However, it is straightforward to implement a custom modification.
For example, useful fields of a modification are 
\field{display_text}, which is displayed in the web interface 
(e.g., {\tt B} for biotin; see \cref{fig:screenshot-initial}),
and
\field{idt_text}, the IDT code for the modification, used for exporting DNA sequences 
(e.g., {\tt "/5Biosg/ACGT"}, which attaches a 5' biotin to the sequence {\tt ACGT}).

Because it is common to attach one type of modification to several strands in a DNA design,
modifications are defined at the top level of a DNA design,
where they are given a string id,
referenced on each strand that contains the modification.

\subsection{scadnano file format}

The following scadnano {\tt .sc} file encodes the design in \cref{fig:screenshot-initial} in a format called JSON, 
a commonly-used plain text format for describing structured data~\cite{json}, with support in many programming language standard libraries.
The format is not exhaustively described here,
but the example shows how the JSON data maps to the data model described above.



\begin{jsonlisting}
{
  "grid": "square",
  "helices": [
    {"max_offset": 48, "grid_position": [0, 0]},
    {"max_offset": 48, "grid_position": [0, 1]}
  ],
  "modifications_in_design": {
    "/5Biosg/": {
      "display_text": "B",
      "idt_text": "/5Biosg/",
      "location": "5'"
    }
  },
  "strands": [
    {
      "color": "#0066cc",
      "sequence": "AACGTAACGTAACGTAACGTAACGTAACGTAACGTAACGTAACGTAACGTAACGTAACGTAACGTAACG",
      "domains": [
        {"helix": 1, "forward": false, "start": 8, "end": 24, "deletions": [20]},
        {"helix": 0, "forward": true, "start": 8, "end": 40, "insertions": [[14,1], [26,2]]},
        {"loopout": 3},
        {"helix": 1, "forward": false, "start": 24, "end": 40}
      ],
      "is_scaffold": true
    },
    {
      "color": "#f74308",
      "sequence": "ACGTTACGTTACGTTTTACGTTACGTTACGTT",
      "domains": [
        {"helix": 1, "forward": true, "start": 8, "end": 24, "deletions": [20]},
        {"helix": 0, "forward": false, "start": 8, "end": 24, "insertions": [[14, 1]]}
      ]
    },
    {
      "color": "#57bb00",
      "sequence": "ACGTTACGTTACGTTACGCGTTACGTTACGTTAC",
      "domains": [
        {"helix": 0, "forward": false, "start": 24, "end": 40, "insertions": [[26, 2]]},
        {"helix": 1, "forward": true, "start": 24, "end": 40}
      ],
      "5prime_modification": "/5Biosg/"
    }
  ]
}
\end{jsonlisting}

\subsection{Comparison to \cadnano\ file format}
\label{sec:file_formats}

The file format used by \cadnano\ v2 is a grid of dimension (number of helices)$\times$(maximum offset) describing at each position whether a domain is present and the direction in which it is going. 
Additional information about \textit{insertions} and \textit{deletions} is given in a similar way. 

An important goal of scadnano is to ensure interoperability with \cadnano\, (see Section~\ref{sec:interop}). 
Thus every cadnano design can be imported into scadnano.
However, the converse is not true; 
scadnano's data model can describe features not present in \cadnano.

\begin{enumerate}
    \item 
    \cadnano\ does not have a way to encode loopouts, modifications, or gridless designs.
    
    \item 
    \cadnano\ does not store DNA sequences in its file format.
    
    \item 
    \cadnano\ has the constraint that helices with even index have the scaffold going forward and helices with odd index have the scaffold going backward. 
    scadnano designs not following that convention cannot be encoded in \cadnano.
    
    \item 
    \label{enum:cadnano-missing-feature-explicit-grid}
    \cadnano\ does not explicitly encode the grid type,
    instead inferring it from the maximum helix offset:
    multiples of 21 represent the honeycomb grid, 
    while multiples of 32 represent the square grid. 
    To encode a scadnano design in \cadnano's convention, each helix's maximum offset is modified to the lowest multiple of 21 or 32 fitting the design.
\end{enumerate}

Converting a scadnano design to \cadnano\ v2 is straightforward: 
lay out all domains of all strands in a (number of helices)$\times$(modified maximum offset) grid. Maximum  offsets have to be modified because of 
\cref{enum:cadnano-missing-feature-explicit-grid}.
However, converting a \cadnano\ design to scadnano format is a bit more involved, 
requiring a connected components detection algorithm performed on the grid---similar to a depth-first search---in order to identify strands and their domains.

\section{Features}
\label{sec:features}

\subsection{Features shared with \cadnano\ v2}
\label{subsec:features:shared_with_cadnano}

The web interface of scadnano is similar to \cadnano\ (see \cref{fig:screenshot-initial}).
Like \cadnano,
scadnano is optimal for structures consisting of parallel helices.
On the left, the side view shows a cross-sectional view of the lattice
where helices can be added to the design. The main view shows what the helix
would look like going from left to right in the screen.
Moving to the
right in the main view
is like moving ``into the screen'' in the side view.

DNA designs are drawn as they are often drawn in figures, with strands on a double-helix represented as straight lines that
are connected to other helices by crossovers.
Users can also add deletions and
insertions (called \emph{skips} and \emph{loops} in \cadnano) which means a strand has fewer or
more bases than the interface's visually depicted length.
Insertions and deletions help to use a regular spacing pattern---note the ``major tick marks'' every 8 bases on the helix---while allowing short regions to deviate and use more or fewer than the typical number of bases between two major tick marks.
One feature scadnano adds to \cadnano\ is the ability to customize the major tick marks, including non-regular spacing, e.g, alternating 10, 11, 10, 11 for single-stranded tiles~\cite{SST2D, drmaurdsa}.

scadnano includes several ``Edit modes'', many similar to those of cadnano, shown in the top right corner of
\cref{fig:screenshot-initial}.
There are two main modes for editing, select mode and pencil mode, as well as several others explained in more detail in the scadnano documentation.
Select mode allows users to select, resize, and
delete items, just like in \cadnano.
(scadnano additionally allows users to copy and paste or move items; see \cref{copy_and_paste}).
Pencil mode is used to
create new objects such as helices, strands, or crossovers.

Users can assign DNA sequences to strands,
and the complementary sequences for the bound
strands are automatically computed. The common M13 DNA sequence is provided as a default
for single-scaffold designs.


\begin{figure}[!htbp]
    \centering
    \begin{subfigure}[b]{0.3\textwidth}
        \includegraphics[width=\textwidth]{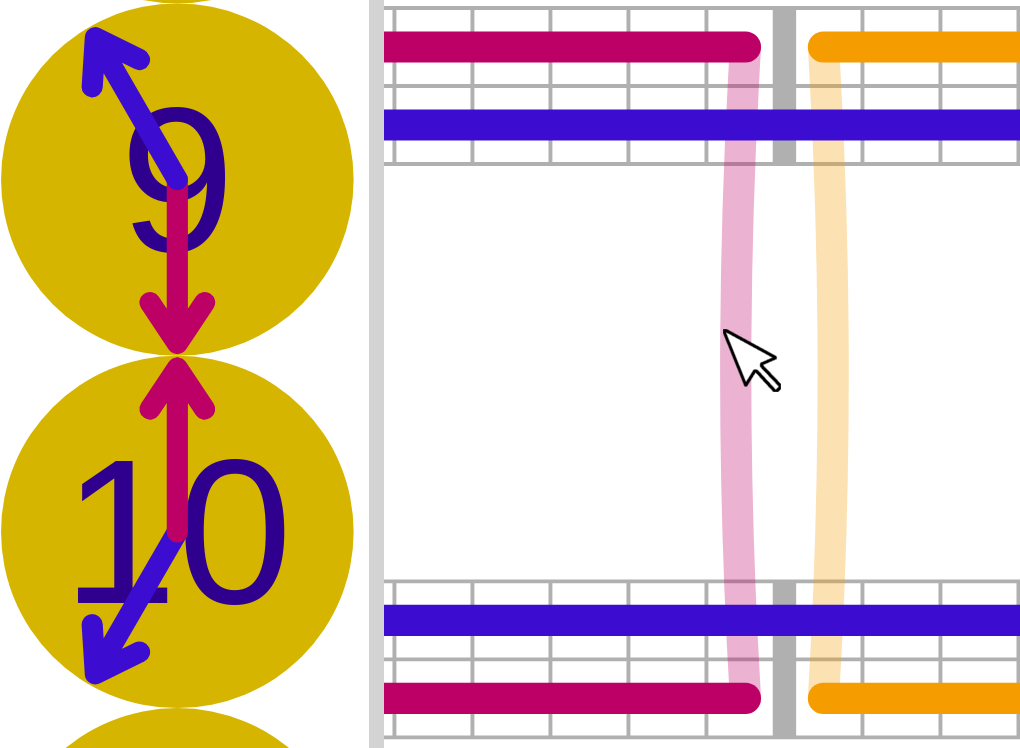}
        \caption{Backbone angles at a crossover.}
        \label{fig:backbone-crossover}
    \end{subfigure}
    \hspace{1cm}
    \begin{subfigure}[b]{0.3\textwidth}
        \includegraphics[width=\textwidth]{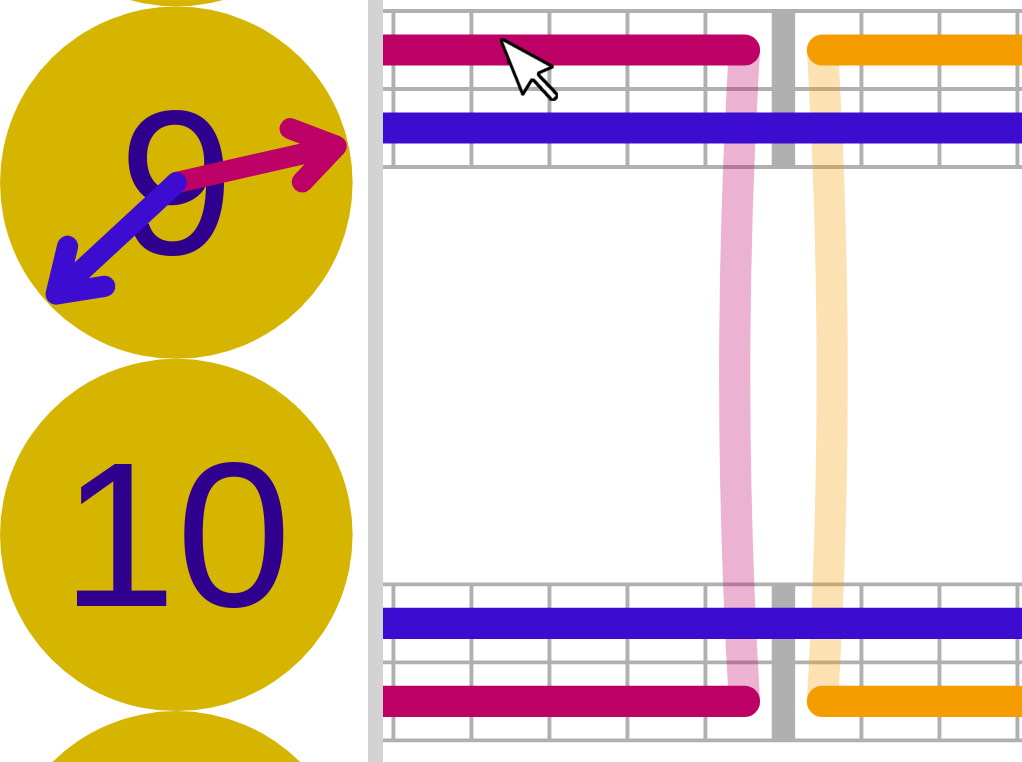}
        \caption{Backbone angle 3 bases to the left.}
        \label{fig:backbone-3-bases-left}
    \end{subfigure}
    \caption{The side view displays the backbone angles to aid with crossover placement.}
    \label{fig:backbone}
\end{figure}

Although scadnano currently provides no 3D visualization,
it does provide a primitive way to visualize the
DNA backbone angles to help pick where to place crossovers; see \cref{fig:backbone}.
This feature
is slightly more flexible than the analogous feature in \cadnano\ in that the user
is allow to set the backbone angle at one base position to see what that
implies about the backbone angle at other (typically nearby) base positions.
For example, a user can ``unstrain'' the backbone at a
crossover so that the backbone angles are perfectly aligned
(see \cref{fig:backbone-crossover}).
The backbone angles at other positions are automatically computed
(see \cref{fig:backbone-3-bases-left}).

The side and main view designs can be exported as SVG figures, and DNA sequences can be be
exported into a CSV file, as well as formats recognized by the synthesis company IDT.


\begin{figure}[!htbp]
    \centering
    \begin{subfigure}[b]{3.5cm}
        \includegraphics[width=3.5cm]{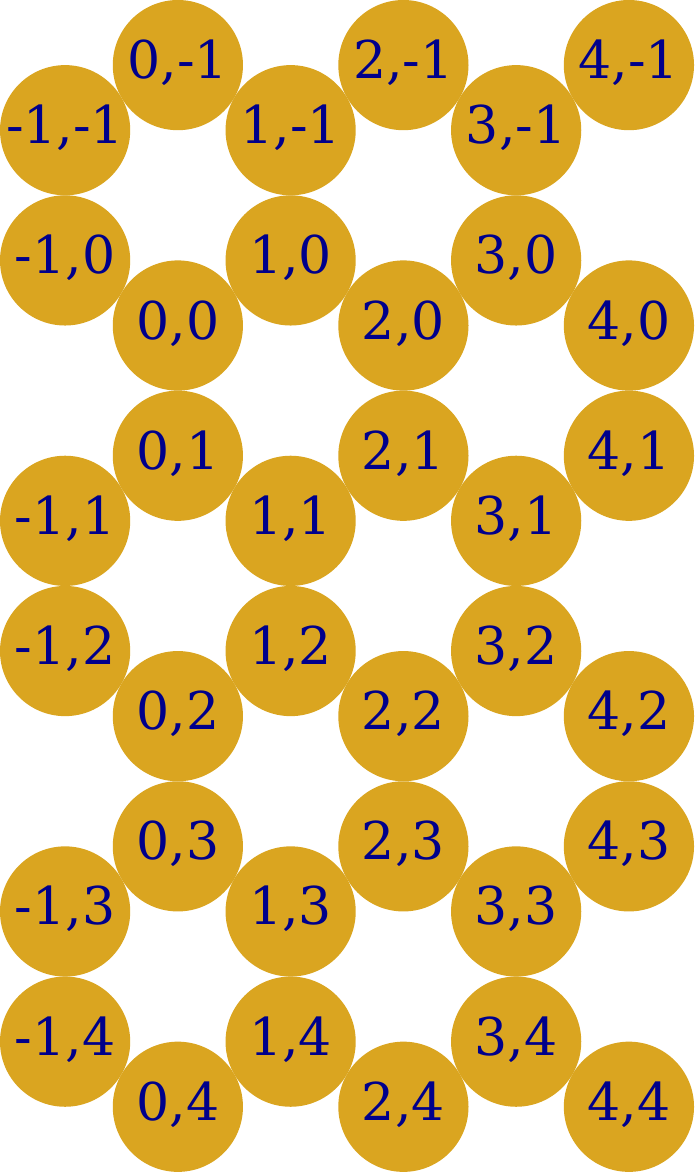}
        \caption{Honeycomb grid, integer coordinates}
        \label{fig:grid_honeycomb}
    \end{subfigure}
    \hspace{0.8cm}
    \begin{subfigure}[b]{4cm}
        \includegraphics[width=4cm]{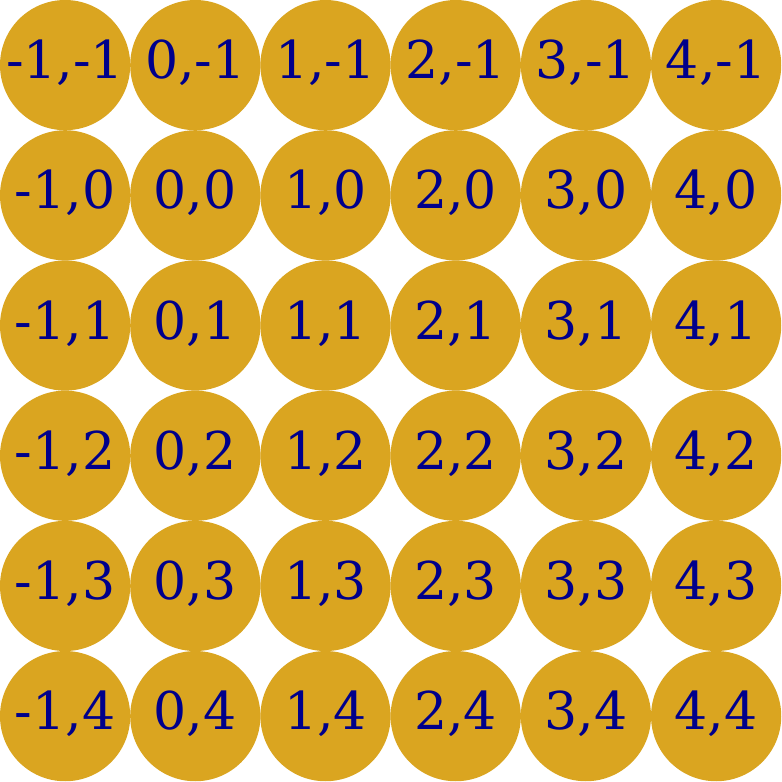}
        \caption{Square grid, integer coordinates}
        \label{fig:grid_square}
    \end{subfigure}
    \hspace{0.8cm}
    \begin{subfigure}[b]{4cm}
        \includegraphics[width=4cm]{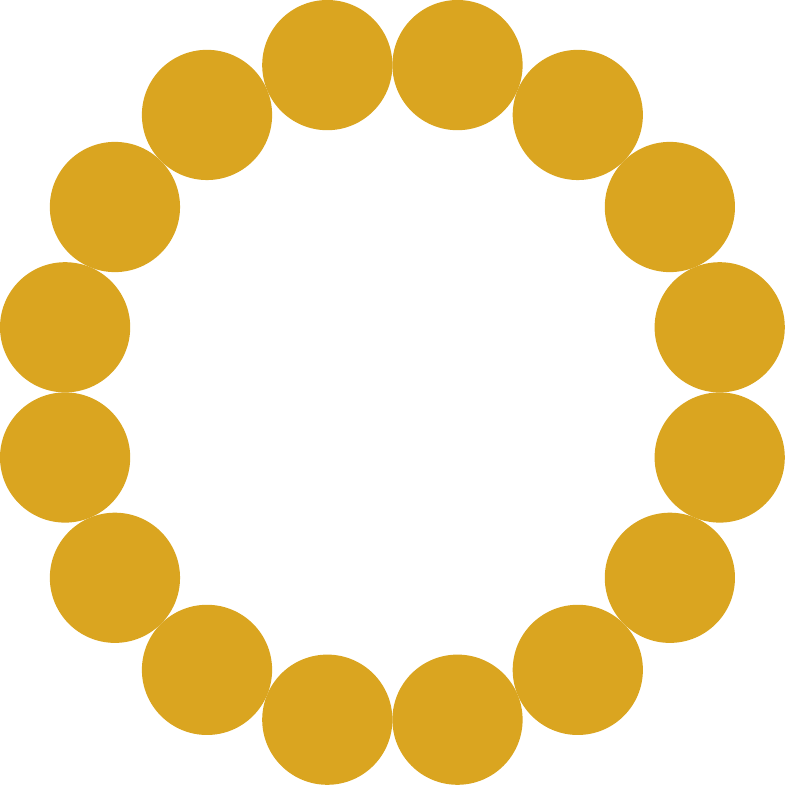}
        \caption{No grid, real-valued coordinates in units of nanometers (coordinates not shown)}
        \label{fig:grid_none}
    \end{subfigure}
    \caption{scadnano grids (hex grid not shown)}
    \label{fig:Grids}
\end{figure}

Like \cadnano, helices can be placed in
a square or honeycomb lattice, as shown in \cref{fig:grid_honeycomb} and~\cref{fig:grid_square}.
scadnano provides two more grids not available on \cadnano:
the hex grid
(allowing helices in the ``holes'' of the honeycomb grid)
and no grid;
see \cref{gridless_helix_placement}.

The remainder of \cref{sec:features} describes features not shared with \cadnano\ v2.

\subsection{Copy and paste}
\label{copy_and_paste}

\begin{figure}[htbp]
    \centering
    \includegraphics[width=\textwidth]{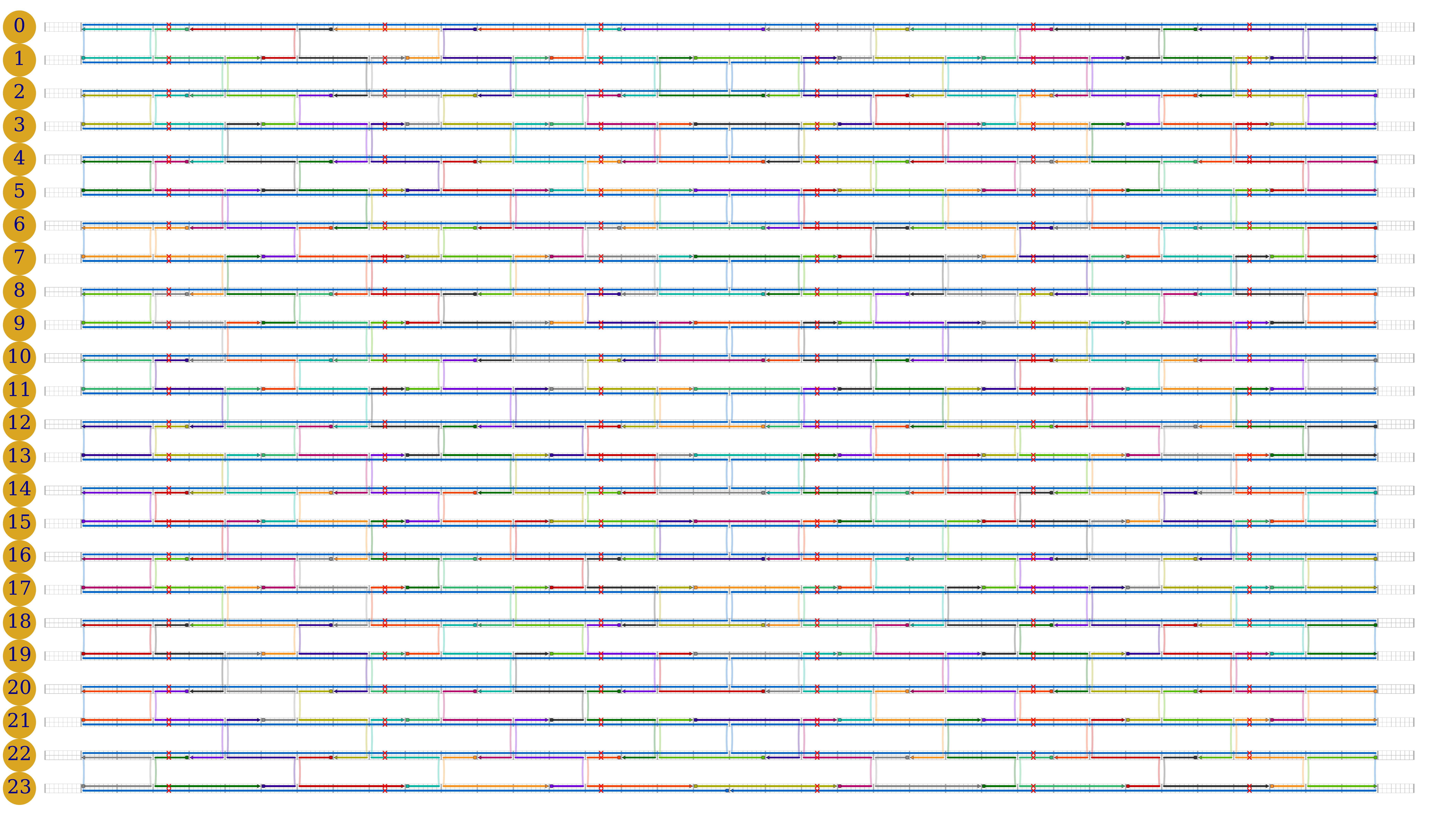}
    \caption{A standard 24 helix DNA origami rectangle design, with ``twist-correction''~\cite{woo2011programmable}.}
    \label{fig:24_helix_rectangle}
\end{figure}

A full DNA origami design using a standard 7249-base M13mp18 scaffold uses $\approx$ 200 staples, which are tedious to create manually.
In scadnano, this process is accelerated by the copy/paste feature.\footnote{
    cadnano provides features to make large designs quickly, \emph{autostaple} and \emph{autobreak}, which are faster than copy/pasting strands, though they give less control over the outcome.
}
For instance, to create a vertical ``column'' of 24 staples in a 24-helix rectangle
(see \cref{fig:24_helix_rectangle}),
one would create 2 types of staples (plus some special cases near the top/bottom), copy/paste them to make 4, copy/paste those to make 8, then copy/paste the group of 8 two more times for a total of 24 staples.
Since most of the design consists of horizontally translated copies of this column,
it can be created quickly by copying and pasting the column.

\subsection{Scripting library}

The scadnano Python module allows one to write scripts for creating and editing scadnano designs.
(Note that cadnano v2.5, unlike v2, does have a scripting library~\cite{cadnano_two_five}, though with incomplete documentation.)
The module helps automate some of the tedious tasks involved in creating DNA designs, as well as making large-scale changes to them that are easier to describe programmatically than to do by hand in scadnano.

For example, the following is Python code generating the design in \cref{fig:24_helix_rectangle},
creating a {\tt .sc} file with the design and a Microsoft Excel file with staple strand DNA sequences in a format ready to order from the DNA synthesis company IDT.
It is perhaps unnecessary to read the code in detail;
we provide it to demonstrate that ``production-ready'' designs can be created with relatively short and simple scripts.
It follows the pattern described in the online tutorial (see first page).

\newpage
\begin{pythonlisting}
import scadnano as sc

def create_design():
    design = create_design_with_precursor_scaffolds()
    add_scaffold_nicks(design)
    add_scaffold_crossovers(design)
    scaffold = design.strands[0]
    scaffold.set_scaffold()
    add_precursor_staples(design)
    add_staple_nicks(design)
    add_staple_crossovers(design)
    add_twist_correcting_deletions(design)
    design.assign_m13_to_scaffold()
    return design

def create_design_with_precursor_scaffolds() -> sc.DNADesign:
    helices = [sc.Helix(max_offset=304) for _ in range(24)]
    scaffolds = [sc.Strand([sc.Domain(helix=helix, forward=helix%2 == 0, start=8, end=296)])
                 for helix in range(24)]
    return DNADesign(helices=helices, strands=scaffolds, grid=square)

def add_scaffold_nicks(design: sc.DNADesign):
    for helix in range(1, 24):
        design.add_nick(helix=helix, offset=152, forward=helix%2 == 0)

def add_scaffold_crossovers(design: sc.DNADesign):
    crossovers = []
    for helix in range(1, 23, 2): # scaffold interior
        crossovers.append(
            sc.Crossover(helix1=helix, helix2=helix+1, offset1=152, forward1=False))
    for helix in range(0, 23, 2): # scaffold edges
        crossovers.append(
            sc.Crossover(helix1=helix, helix2=helix+1, offset1=8, forward1=True, half=True))
        crossovers.append(
            sc.Crossover(helix1=helix, helix2=helix+1, offset1=295, forward1=True,half=True))
    design.add_crossovers(crossovers)

def add_precursor_staples(design: sc.DNADesign):
    staples = [sc.Strand([sc.Domain(helix=helix, forward=helix%2 == 1, start=8, end=296)])
               for helix in range(24)]
    for staple in staples:
        design.add_strand(staple)

def add_staple_nicks(design: sc.DNADesign):
    for helix in range(24):
        start_offset = 32 if helix % 2 == 0 else 48
        for offset in range(start_offset, 280, 32):
            design.add_nick(helix, offset, forward=helix%2 == 1)

def add_staple_crossovers(design: sc.DNADesign):
    for helix in range(23):
        start_offset = 24 if helix % 2 == 0 else 40
        for offset in range(start_offset, 296, 32):
            if offset != 152:  # skip crossover near seam
                design.add_full_crossover(helix1=helix, helix2=helix + 1,
                                          offset1=offset, forward1=helix % 2 == 1)

def add_twist_correcting_deletions(design: sc.DNADesign):
    for helix in range(24):
        for offset in range(27, 294, 48):
            design.add_deletion(helix, offset)

def export_idt_plate_file(design: sc.DNADesign):
    for strand in design.strands:
        if strand != design.scaffold:
            strand.set_default_idt(use_default_idt=True)
    design.write_idt_plate_excel_file(use_default_plates=True)

if __name__ == "__main__":
    design = create_design()
    export_idt_plate_file(design)
    design.write_scadnano_file()
\end{pythonlisting}



\subsection{Hiding helices to aid 3D design}

\begin{figure}[!htbp]
    \centering
    \begin{subfigure}[t]{0.35\textwidth}
        \includegraphics[width=\textwidth]{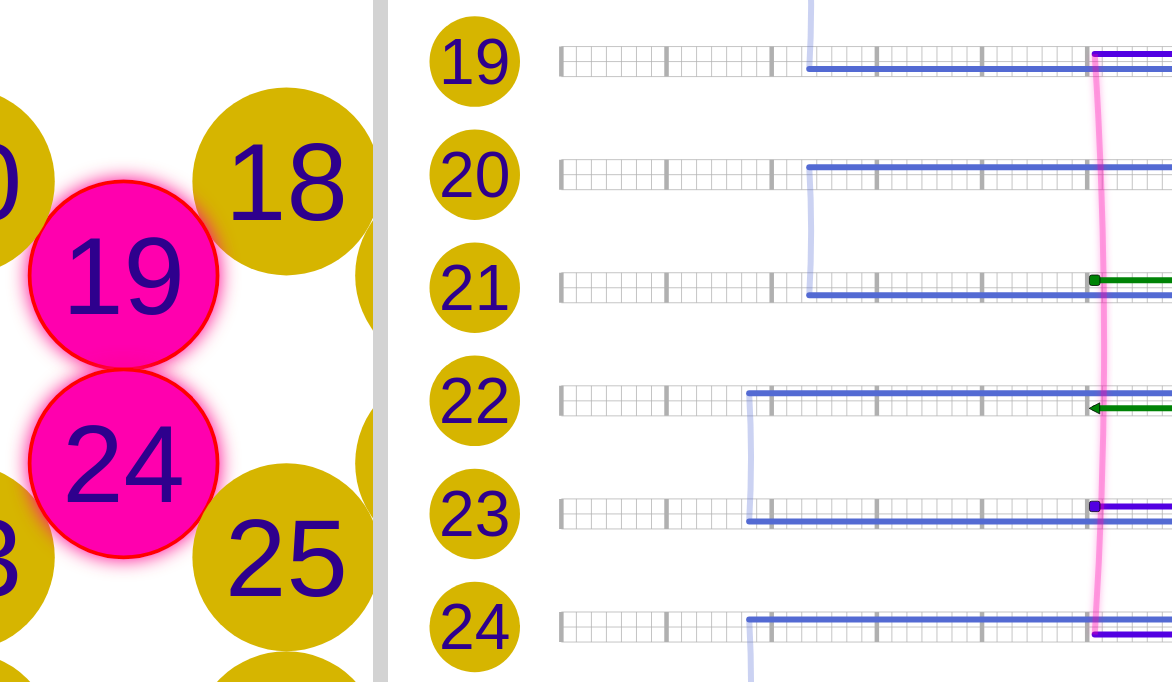}
        \caption{Without helix-hiding}
        \label{fig:squarenut_design-helices-19-and-24}
    \end{subfigure}
    \hspace{1cm}
    \begin{subfigure}[t]{0.4\textwidth}
        \includegraphics[width=\textwidth]{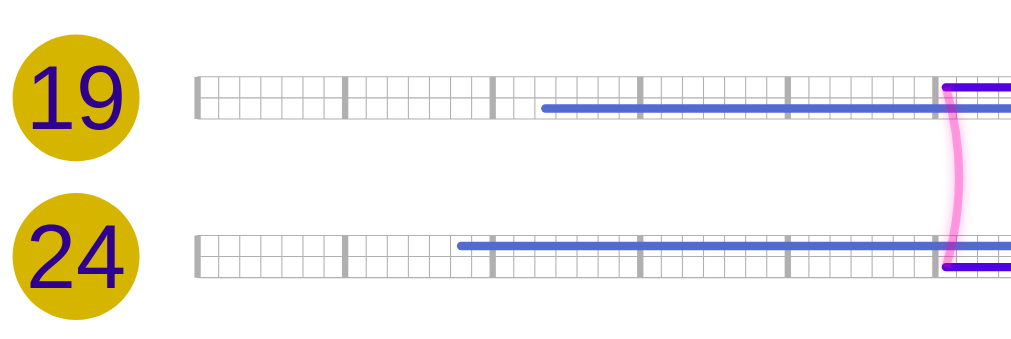}
        \caption{With helix-hiding}
        \label{fig:squarenut_design-helices-19-and-24_with_helix_hide}
    \end{subfigure}

    \caption{
    Two helices in a design, 19 and 24, are adjacent in the side view (i.e., in the actual 3D structure) but not in the main view.
    The selected crossover appears ``long-range'' in
    \cref{fig:squarenut_design-helices-19-and-24}, but ``short-range'' in \cref{fig:squarenut_design-helices-19-and-24_with_helix_hide}.
}
    \label{fig:square_nut_19_24}
\end{figure}

The 2D main view in scadnano
distorts the relative positions of the helices if they do not form a flat 2D shape as in \cref{fig:24_helix_rectangle}.
For example, consider \cref{fig:square_nut_19_24}.
Helices 19 and 24, though adjacent (see side view),
appear far apart in the main view.
Thus crossovers between these helices, while appearing to stretch over a long distance (\cref{fig:squarenut_design-helices-19-and-24}),
are the same length as any other crossover
(just a single phosphate group between two DNA bases).

\begin{figure}[th]
    \centering
    \begin{subfigure}[t]{\textwidth}
        \includegraphics[width=\textwidth]{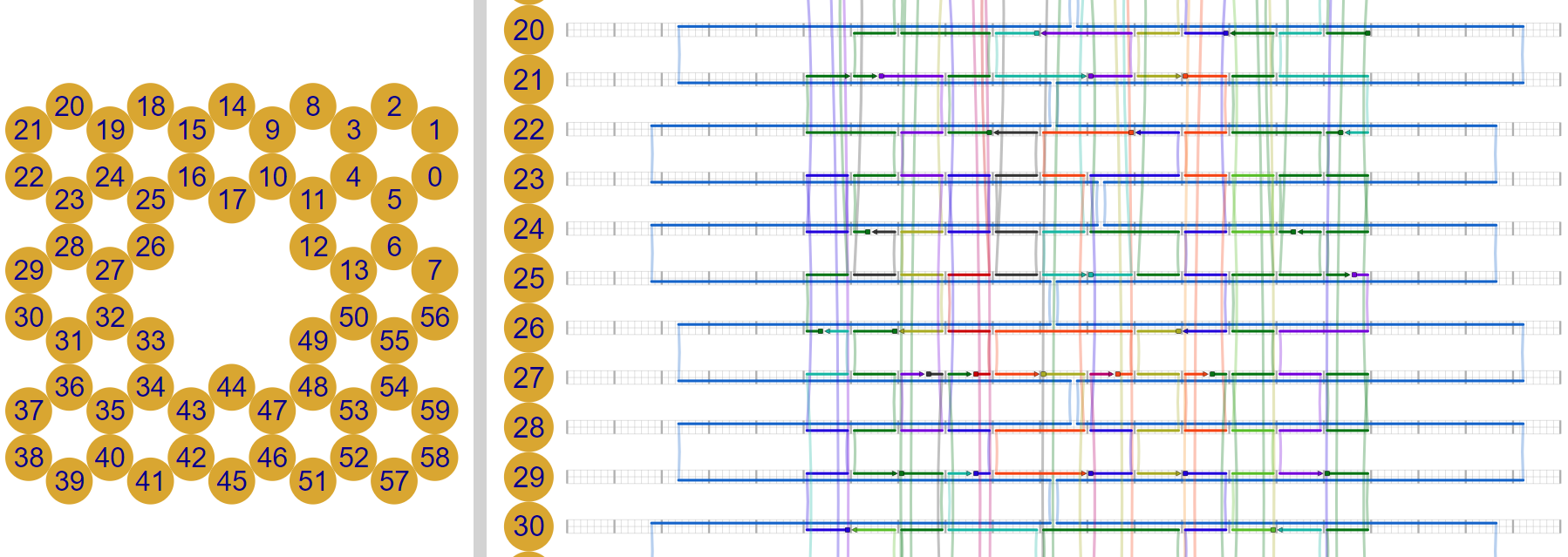}
        \caption{All helices shown, causing the dreaded \emph{crossover cobweb}, like laser beams guarding priceless art.}
        \label{fig:squarenut_not_selected}
    \end{subfigure}

    \begin{subfigure}[t]{\textwidth}
        \includegraphics[width=\textwidth]{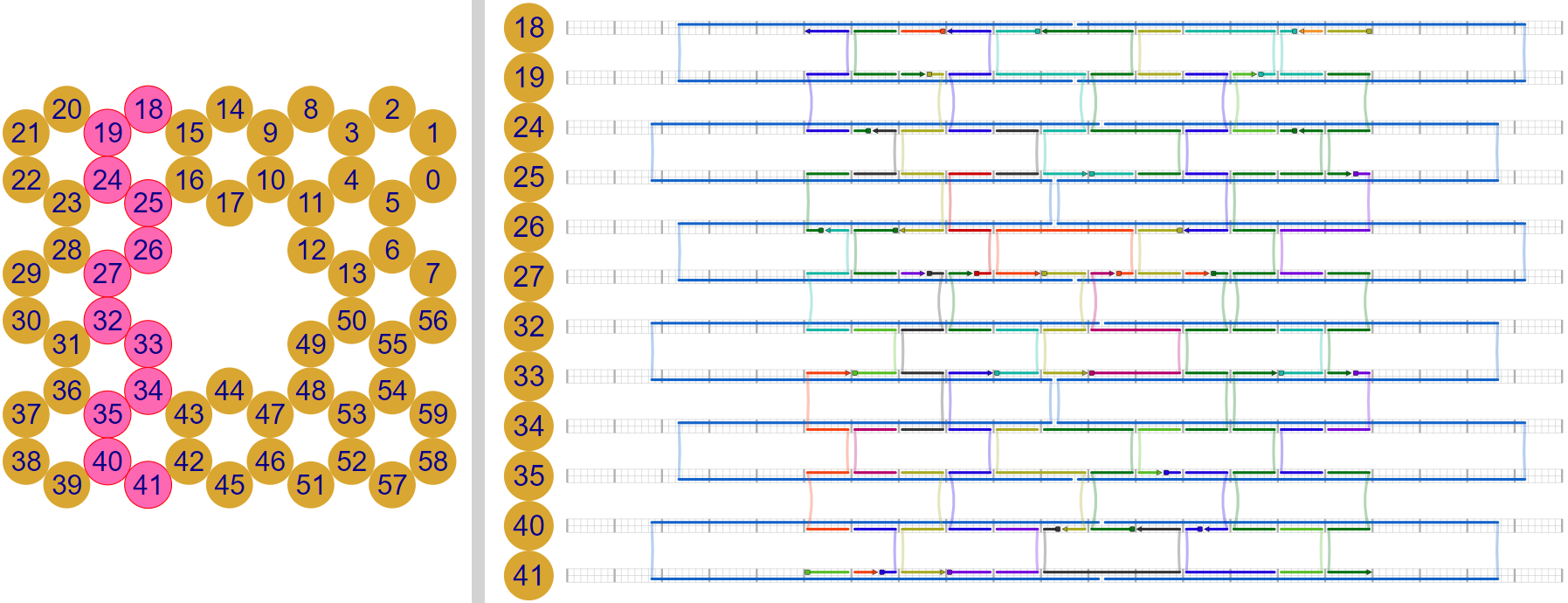}
        \caption{Restricted subset of helices displayed: only relevant helices and crossovers are shown.}
        \label{fig:squarenut_selected}
    \end{subfigure}
    \caption{Squarenut 3D origami~\cite{douglas2009self}, a typical 3D origami difficult to visualize in a 2D projection.}
    \label{fig:square_nut}
\end{figure}

This can make it difficult to analyze and edit 3D designs.
For example, consider the squarenut design from the original 3D origami paper~\cite{douglas2009self}
(see \cref{fig:squarenut_not_selected}).
This design is difficult to visualize because the 2D view is not representative of the 3D positions of the actual DNA helices,
in no small part because of the ``cobweb'' of crossovers that results.

To aid in visualization,
scadnano can display only selected helices
(see \cref{fig:squarenut_selected}).
Helix 19 and 24 in \cref{fig:squarenut_design-helices-19-and-24_with_helix_hide}
can be seen in the side view are actually adjacent in 3D space.
When other helices are hidden, helices 19 and 24 are displayed adjacently in the main view.

\cadnano\ puts all helices immediately adjacent to each other in the order they are displayed in the main view.
scadnano uses the distance between helices
(as determined by their grid position or gridless 3D position)
to determine distances.
Helices are displayed in order of their index field \field{idx} (unless \field{helices\_view\_order} is specified to alter this order),
but two helices adjacent in this order will have a vertical distance between them in the main view proportional to the distance as determined by the grid position or gridless 3D position.

\subsection{Single-stranded loopouts}

scadnano allows a type of single-stranded domain not associated to any helix, called a \emph{loopout},
used to describe common single-stranded features such as hairpins.
In \cadnano\, users would need to make a ``fake'' helix if they want to add a single-stranded DNA.
For some designs, this creates awkward artifacts such as long-range crossovers to reach the fake helix.

\subsection{DNA modifications}

\def\afmwidth{4.4in}

\begin{figure}[h]
     \centering
    \begin{subfigure}[b]{0.93\textwidth}
        \hspace{0.2cm}
        \includegraphics[width=0.95\textwidth]{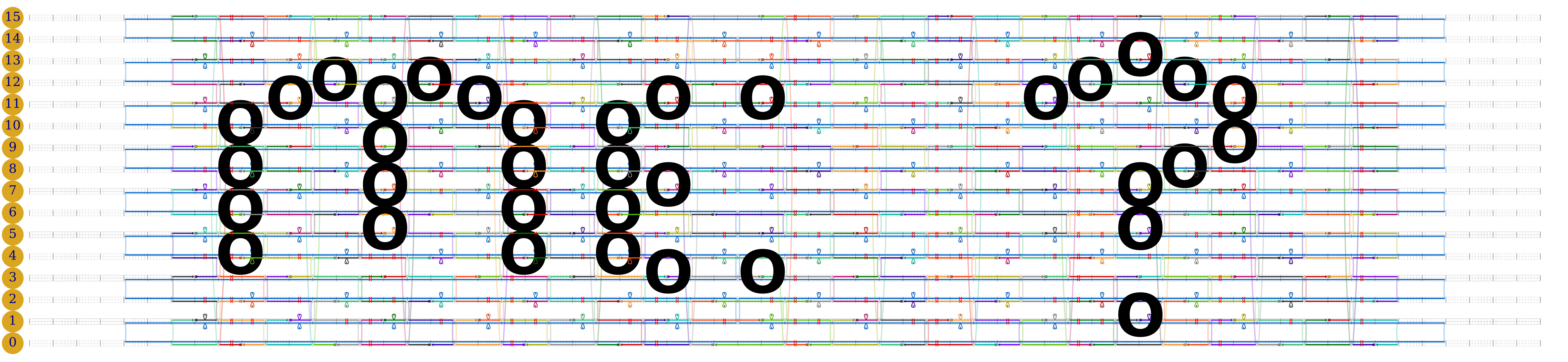}
        \caption{biotin DNA modifications on the 5' end of some staples, displayed in scadnano.}
        \label{fig:me_main}
    \end{subfigure}
    \begin{subfigure}[b]{\afmwidth}
        \begin{tikzpicture} 
            \node[anchor=south west,inner sep=0] at (0,0) {\includegraphics[width=\afmwidth]{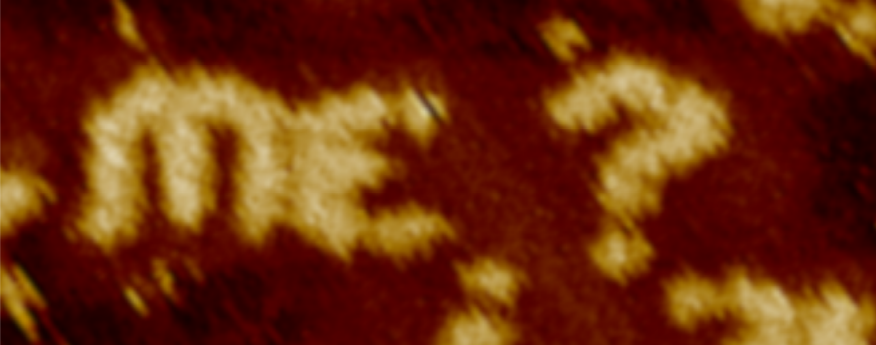}};
            \fill[white] (1.0, 0.3) rectangle (4.6, 0.5);
        \end{tikzpicture}
        \caption{The same design imaged with atomic force microscopy (AFM), with streptavidin added to visualize the biotin locations.
        (scale bar: 50 nm)
        (image source:\hspace{1cm} \url{https://web.cs.ucdavis.edu/~doty/papers/\#proposal}) 
        }
        \label{fig:me_dna}
    \end{subfigure}
    \caption{An example of a design containing biotin modifications.}
    \label{fig:modification}
\end{figure}

scadnano supports for DNA modifications, such as biotin or Cy3~\cite{idt_mods}.
\cref{fig:me_main} shows an example of biotin modifications to the 5' end of some staples in a
16-helix DNA origami.
Users can specify
a string such as \lstinline{"O"} to represent the modification in the web interface.

The aspect ratio is proper for 2D origami with helices all stacked in the square lattice, helping to place modifications and visualize their relative positions to scale.
Compare the scadnano display in \cref{fig:me_main} to the AFM image in \cref{fig:me_dna}.
Currently, only a few pre-loaded modifications are provided,
but users can describe custom modifications.

\subsection{Unused fields}
\label{sec:unused_fields}
In order to maximize interoperability with other tools,
scadnano allows arbitrary fields to be included in a scadnano {\tt .sc} file.
Any fields that it does not recognize are simply ignored.
However, they are stored and written back out when the file is saved.
Thus, ``light'' editing of scadnano files is possible that will preserve fields used by other programs.
For example, codenano~\cite{codenano} allows an optional field \field{label} on each strand,
which will be preserved for each strand by scadnano while editing other aspects of the design.

\subsection{Gridless helix placement}
\label{gridless_helix_placement}

scadnano includes the option to use no grid;
see \cref{fig:grid_none}.
This allows more flexible helix placement, where helix centers can be placed at any real-valued (i.e., floating-point) $(x,y)$ coordinate.
This feature is useful for some
designs that do not align nicely with the standard square or honeycomb lattice.
In the absence of a grid, coordinates of helices are specified in nanometers.
By default, the distance between each DNA helix center is 2.5 nm.\footnote{
    The accepted measurement of the DNA double-helix diameter is $\approx$ 2 nm.
    However, AFM images show that in 2D square-lattice DNA origami designs,
    an origami with $n$ helices will have height in nanometers of approximately $2.5 \cdot n$ due to electrostatic repulsion between neighboring helices.
}

\subsection{Interoperability with \cadnano}
\label{sec:interop}


Interoperability with \cadnano\ (version 2) is an important goal of the project. Both the scadnano GUI and Python module provide functionality that allows users to import/export a design from/to \cadnano.
All \cadnano\ (version 2) designs can be imported in scadnano.
However, because of fundamental differences between the way \cadnano\ and scadnano encode designs,
some scadnano designs cannot be converted \cadnano\ (see \cref{sec:file_formats}).\footnote{
    These constraints are described in the documentation:
    \scriptsize{\url{https://scadnano-python-package.readthedocs.io/en/latest/index.html\#interoperability-cadnano-v2}}
}

\section{Software architecture}
\label{sec:architecture}

\subsection{Two codebases}
The codebase for scadnano is split into two pieces:
the Python scripting library,
and the web interface.
Unfortunately, some algorithmic functionality is duplicated between them.
We chose Python as the scripting language because it is easy to learn and already familiar to many physical scientists likely to use scadnano.
However 
(despite innovations such as 
Pyodide~\cite{pyodide}, 
Skulpt~\cite{skulpt},
and Brython~\cite{brython}),
Python is not well-suited for
\emph{front-end} web programming,
where the code is executed in the browser rather than on a server.
A design goal of scadnano is to do as much work as possible in the browser.

The web interface is instead implemented using the Dart programming language~\cite{dart},
a modern, strongly-typed, object-oriented language that can be compiled to Javascript, 
the \emph{lingua franca} of web browsers.
In order to make the Python scripting library as easy to use as possible
(no dependence on Dart libraries)
and to keep the web interface as fast as possible and avoid the need to farm out computation to a server,
some algorithms (e.g., computing complementary DNA sequences of strands when they are bound to another strand that has had a DNA sequence assigned to it)
are implemented in both libraries.

However, we intend for the file format to be decoupled from the scripting and web-based programs that manipulate it.
Indeed, another tool called codenano~\cite{codenano} uses essentially the same file format as scadnano,
although that program is written in Rust and has the user specify the design by writing Rust code.

\subsection{Unidirectional data flow in graphical user interface code}
Graphical user interface software, inherently asynchronous and non-sequential, is notoriously difficult to reason about.
Whole classes of bugs exist that do not plague programs with only sequential logic.
The open-source software community has developed many tools to aid in such design.
The \emph{model-view-controller} (MVC) architecture is almost as old as graphical interfaces themselves,
dating to the 1970s~\cite{pope1988cookbook}.
However, MVC is not very well-defined, particularly the controller part,
and still lends itself to common bugs.

A more recent innovation,
originating within the past decade,
goes under a few names, such 
as 
\emph{model-view-update}, 
\emph{the Elm architecture}~\cite{elm},
or 
\emph{unidirectional data flow}~\cite{redux_unidirectional}.
Several variants exist implementing the idea.
We chose a popular pair of technologies,
React~\cite{react} and Redux~\cite{redux}.
They are designed for Javascript,
but since Dart compiles to Javascript,
they can be used with Dart with appropriate wrapping libraries~\cite{overreact, redux_dart}.

The cited links go into detail about the architecture;
we summarize it briefly here for the curious.
Briefly,
all application \emph{state} is stored in a single immutable object.
(In scadnano, this includes the entire DNA design, as well as more ephemeral UI state, such as which strands are currently selected.)
Immutability is a powerful concept in programming,
allowing one to share an object between many concurrent processes without worrying that one process will modify it in ways unexpected by the other processes.
The global state object is a tree (cycles are difficult to handle with immutable objects).
The \emph{view} (what the user sees on the screen) is specified as a deterministic function of the state.
This greatly reduces the ``surface area'' where bugs can (and reliably do) occur:
the application does not have to contain code stating how to \emph{modify} the view in response to any possible change in the state.
It merely says what the \emph{entire} view should be, as a function of the \emph{entire} state.

Changes to the application state are expressed using the Command pattern~\cite{gamma1995design} by dispatching an \emph{action} describing that the state should change.
The application responds to the action by computing the new state as a deterministic function of the old state and the action.
The view redraws itself, but optimizations ensure only the parts that depend on changed state will actually be redrawn.

This decoupling of actions that change state (and the sometimes complex logic behind them),
and views that draw themselves as a function of a single state, 
is the key to making it straightforward to implement new features without introducing bugs.
It's not foolproof; bugs do occur.
There is also a nontrivial computational cost:
the React library compares the old state to the new to determine which subtrees actually changed (determining which parts of the view actually need to re-render), a potentially expensive operation.

However, we find it is worth the computational cost for the benefit of reliability.
We believe it will make it easier to maintain scadnano, fix bugs, and add features in the future.

Both the Python package and the Dart web interface are open-source software to which anyone can contribute. 
Both repositories have a CONTRIBUTING document explaining how to contribute to the projects, following the git model of making a separate branch, adding the change, and doing a pull request to merge the changes. 
Both repositories are currently maintained by the first author, who reviews all pull requests.

\section{Conclusion}
\label{sec:conclusion}

The goal of scadnano is to reproduce the usefulness of \cadnano\ for designing large-scale DNA structures in a web app with a well-documented, easy-to-use scripting library.
It is ready to use for designing DNA structures, although
some work remains to bring it up to a more polished state.
The issues page of each repository (see first page) shows many bugs and feature enhancements that have not yet been addressed.

scadnano excels where \cadnano\ excels:
in describing DNA structures where all DNA helices are in parallel.
A broader range of DNA nanostructures exists,
such as wireframe designs~\cite{benson2015dna, zhang2015complex}
and curved DNA origami shapes~\cite{dietz2009folding, han2011dna}.
A 2D projected view can describe these, 
but more awkwardly than a 3D view.
Since the chief goal of scadnano is to remain easy to use and responsive to bug reports and feature requests within the current scope of scadnano,
it will remain for the near-term future as a tool primarily for designs that are straightforward to visualize in 2D.
We outline possible future work:

\begin{description}

\item[export to other file formats.]
    Currently, scadnano can export to the \cadnano\ v2 file format,
    and it can export DNA sequences in either a comma-separated value (CSV) file, 
    which can be processed by the user's custom scripts,
    or in a few formats recognized by the DNA synthesis company IDT (Integrated DNA Technologies, Coralville, IA, \url{https://www.idtdna.com}).
    It should be straightforward to export to formats recognized by other DNA synthesis companies (e.g., Bioneer),
    or other DNA nanotech software (e.g., oxDNA).
    
\item[helices rotated in the main view plane.]
    Some 2D structures do not have all helices in parallel,
    for example DNA origami implementations of 4-sided tiles~\cite{tikhomirov2017programmable}, 
    or
    flat origami ``stiffened'' by a second layer of perpendicular helices~\cite{thubagere2017cargo}.
    We are exploring design ideas for supporting this in a way ``natural'' for editing in the 2D view.
    In particular, copy/paste and moving of strands spanning multiple helices makes most sense for groups of helices that are parallel.
    One idea is to let a design specify several helix \emph{groups}, where all helices within a group are parallel,
    but the groups have different rotations and translations.
    (For example, there would be two groups for~\cite{thubagere2017cargo} and two or four groups for~\cite{tikhomirov2017programmable}.)
    
\item[3D visualization.]
    \cadnano\ has never been ideal for visualizing arbitrary 3D structures, and neither is scadnano currently.
    It may remain the case that the ideal way to visualize 3D structures is to export the design to another tool specialized for the job,
    such as 
    codenano~\cite{codenano}, 
    CanDo~\cite{cando}, 
    or oxDNA~\cite{snodin2015introducing}.
    However,
    WebGL provides a powerful platform for visualizing 3D structures, used by other software such as oxDNA and codenano.
    In fact, since codenano is itself implemented as a web app (written in Rust that is compiled to WebAssembly, which is itself callable from Javascript),
    it should be possible to implement the 3D visualization features of codenano as a library that scadnano can call.
    
\item[DNA design database.]
    Communication of DNA designs through the Supplementary Information of a journal remains an ad hoc method.
    A centralized database of DNA designs would benefit the community.
    We hope that the scadnano/codenano file format is sufficiently expressive to describe any such design.
    However, such a database need not have anything to do with the scadnano website itself.
    
\item[collaborative editing.]
    Collaborative editing tools such as Google Docs make use of a recently developed technique known as a \emph{conflict-free replicated data type} (CRDT)~\cite{shapiro2011conflict}.
    It is conceivable that a CRDT representation of a DNA design could enable remote collaborators to simultaneously view and edit a DNA design.
\end{description}

\bibliography{ref}

\begin{thebibliography}{10}

\bibitem{brython}
Brython.
\newblock \url{https://brython.info/}.

\bibitem{cadnano_two_five}
\cadnano\ v2.5.
\newblock \url{https://github.com/cadnano/cadnano2.5}.

\bibitem{cadnano_two_five_api}
\cadnano\ v2.5 {P}ython {API}.
\newblock \url{https://cadnano.readthedocs.io/en/master/scripting.html}.

\bibitem{cando}
Cando.
\newblock \url{https://cando-dna-origami.org/}.

\bibitem{codenano}
codenano.
\newblock \url{https://dna.hamilton.ie/2019-07-18-codenano.html}.

\bibitem{dart}
Dart programming language.
\newblock \url{https://dart.dev/}.

\bibitem{elm}
Elm programming language.
\newblock \url{https://elm-lang.org/}.

\bibitem{idt_mods}
{IDT} {DNA} modifications.
\newblock
  \url{https://www.idtdna.com/pages/products/custom-dna-rna/oligo-modifications}.

\bibitem{json}
Json (javascript object notation).
\newblock \url{https://www.json.org/json-en.html}.

\bibitem{overreact}
Overreact {D}art library.
\newblock \url{https://pub.dev/packages/over_react}.

\bibitem{pyodide}
Pyodide.
\newblock \url{https://github.com/iodide-project/pyodide}.

\bibitem{react}
React {J}avascript library.
\newblock \url{https://reactjs.org/}.

\bibitem{redux_dart}
Redux {D}art library.
\newblock \url{https://pub.dev/packages/redux}.

\bibitem{redux}
Redux {J}avascript library.
\newblock \url{https://redux.js.org/}.

\bibitem{skulpt}
Skulpt.
\newblock \url{https://skulpt.org/}.

\bibitem{redux_unidirectional}
Unidirectional data flow in {R}edux.
\newblock \url{https://redux.js.org/basics/data-flow}.

\bibitem{samson}
{SAMSON}, the open molecular modeling platform.
\newblock \url{ https://www.samson-connect.net}, 2019.

\bibitem{vhelix}
Erik Benson, Abdulmelik Mohammed, Johan Gardell, Sergej Masich, Eugen Czeizler,
  Pekka Orponen, and Bj\"{o}rn H\"{o}gberg.
\newblock {DNA} rendering of polyhedral meshes at the nanoscale.
\newblock {\em Nature}, 523(7561):441--444, July 2015.
\newblock \href {https://doi.org/10.1038/nature14586}
  {\path{doi:10.1038/nature14586}}.

\bibitem{benson2015dna}
Erik Benson, Abdulmelik Mohammed, Johan Gardell, Sergej Masich, Eugen Czeizler,
  Pekka Orponen, and Bj{\"o}rn H{\"o}gberg.
\newblock {DNA} rendering of polyhedral meshes at the nanoscale.
\newblock {\em Nature}, 523(7561):441--444, 2015.

\bibitem{chatterjee2017spatially}
Gourab Chatterjee, Neil Dalchau, Richard~A Muscat, Andrew Phillips, and Georg
  Seelig.
\newblock A spatially localized architecture for fast and modular {DNA}
  computing.
\newblock {\em Nature nanotechnology}, 12(9):920, 2017.

\bibitem{adenita}
Elisa de~Llano, Haichao Miao, Yasaman Ahmadi, Amanda~J. Wilson, Morgan Beeby,
  Ivan Viola, and Ivan Barisic.
\newblock Adenita: Interactive 3{D} modeling and visualization of {DNA}
  nanostructures.
\newblock Technical report, bioRxiv, 2019.
\newblock URL: \url{https://doi.org/10.1101/849976}.

\bibitem{dietz2009folding}
Hendrik Dietz, Shawn~M Douglas, and William~M Shih.
\newblock Folding {DNA} into twisted and curved nanoscale shapes.
\newblock {\em Science}, 325(5941):725--730, 2009.

\bibitem{douglas2009self}
Shawn~M Douglas, Hendrik Dietz, Tim Liedl, Bj{\"o}rn H{\"o}gberg, Franziska
  Graf, and William~M Shih.
\newblock Self-assembly of {DNA} into nanoscale three-dimensional shapes.
\newblock {\em Nature}, 459(7245):414--418, 2009.

\bibitem{douglas2009rapid}
Shawn~M Douglas, Adam~H Marblestone, Surat Teerapittayanon, Alejandro Vazquez,
  George~M Church, and William~M Shih.
\newblock Rapid prototyping of 3{D} {DNA}-origami shapes with ca{DNA}no.
\newblock {\em Nucleic Acids Research}, 37(15):5001--5006, 2009.
\newblock \url{https://cadnano.org/}.

\bibitem{gamma1995design}
Erich Gamma, Richard Helm, Ralph Johnson, and John Vlissides.
\newblock {\em Design patterns: Elements of reusable object-oriented software}.
\newblock Pearson Education India, 1995.

\bibitem{gu2010proximity}
Hongzhou Gu, Jie Chao, Shou-Jun Xiao, and Nadrian~C Seeman.
\newblock A proximity-based programmable {DNA} nanoscale assembly line.
\newblock {\em Nature}, 465(7295):202--205, 2010.

\bibitem{han2011dna}
Dongran Han, Suchetan Pal, Jeanette Nangreave, Zhengtao Deng, Yan Liu, and Hao
  Yan.
\newblock {DNA} origami with complex curvatures in three-dimensional space.
\newblock {\em Science}, 332(6027):342--346, 2011.

\bibitem{athena}
Hyungmin Jun, Xiao Wang, William Bricker, Steve Jackson, and Mark Bathe.
\newblock Rapid prototyping of wireframe scaffolded {DNA} origami using
  {ATHENA}.
\newblock Technical report, bioRxiv, 2020.
\newblock \href {https://doi.org/10.1101/2020.02.09.940320}
  {\path{doi:10.1101/2020.02.09.940320}}.

\bibitem{pope1988cookbook}
Glenn Krasner and Stephen Pope.
\newblock A cookbook for using the model-view-controller user interface
  paradigm in {S}malltalk-80.
\newblock {\em Journal of object-oriented programming}, 1, 1988.

\bibitem{viennaRNA}
Ronny Lorenz, Stephan~H Bernhart, Christian~H\"{o}ner zu~Siederdissen, Hakim
  Tafer, Christoph Flamm, Peter~F Stadler, and Ivo~L Hofacker.
\newblock {ViennaRNA} package 2.0.
\newblock {\em Algorithms for Molecular Biology}, 6(1), November 2011.
\newblock \href {https://doi.org/10.1186/1748-7188-6-26}
  {\path{doi:10.1186/1748-7188-6-26}}.

\bibitem{mrdna}
Christopher Maffeo and Aleksei Aksimentiev.
\newblock Mr{DNA}: A multi-resolution model for predicting the structure and
  dynamics of nanoscale dna objects.
\newblock {\em bioRxiv}, 2019.
\newblock URL: \url{https://www.biorxiv.org/content/early/2019/12/05/865733},
  \href {https://doi.org/10.1101/865733} {\path{doi:10.1101/865733}}.

\bibitem{crisscross_assembly}
Dionis Minev, Christopher~M. Wintersinger, Anastasia Ershova, and William~M
  Shih.
\newblock Robust nucleation control via crisscross polymerization of {DNA}
  slats.
\newblock Technical report, biorXiv, 2019.
\newblock URL:
  \url{https://www.biorxiv.org/content/10.1101/2019.12.11.873349v1}.

\bibitem{rothemund2006folding}
Paul W.~K. Rothemund.
\newblock Folding {DNA} to create nanoscale shapes and patterns.
\newblock {\em Nature}, 440(7082):297--302, 2006.

\bibitem{shapiro2011conflict}
Marc Shapiro, Nuno Pregui{\c{c}}a, Carlos Baquero, and Marek Zawirski.
\newblock Conflict-free replicated data types.
\newblock In {\em SSS 2011: Symposium on self-stabilizing systems}, pages
  386--400, 2011.

\bibitem{snodin2015introducing}
Benedict~EK Snodin, Ferdinando Randisi, Majid Mosayebi, Petr {\v{S}}ulc, John~S
  Schreck, Flavio Romano, Thomas~E Ouldridge, Roman Tsukanov, Eyal Nir, Ard~A
  Louis, and Jonathan P.~K. Doye.
\newblock Introducing improved structural properties and salt dependence into a
  coarse-grained model of {DNA}.
\newblock {\em The Journal of chemical physics}, 142(23):234901, 2015.

\bibitem{thubagere2017cargo}
Anupama~J Thubagere, Wei Li, Robert~F Johnson, Zibo Chen, Shayan Doroudi,
  Yae~Lim Lee, Gregory Izatt, Sarah Wittman, Niranjan Srinivas, Damien Woods,
  Erik Winfree, and Lulu Qian.
\newblock A cargo-sorting {DNA} robot.
\newblock {\em Science}, 357(6356):eaan6558, 2017.

\bibitem{tikhomirov2017programmable}
Grigory Tikhomirov, Philip Petersen, and Lulu Qian.
\newblock Programmable disorder in random {DNA} tilings.
\newblock {\em Nature nanotechnology}, 12(3):251, 2017.

\bibitem{oxDNA}
Petr \v{S}ulc, Flavio Romano, Thomas~E. Ouldridge, Lorenzo Rovigatti, Jonathan
  P.~K. Doye, and Ard~A. Louis.
\newblock Sequence-dependent thermodynamics of a coarse-grained {DNA} model.
\newblock {\em The Journal of Chemical Physics}, 137(13):135101, 2012.
\newblock URL: \url{http://link.aip.org/link/?JCP/137/135101/1}, \href
  {https://doi.org/10.1063/1.4754132} {\path{doi:10.1063/1.4754132}}.

\bibitem{SST2D}
Bryan Wei, Mingjie Dai, and Peng Yin.
\newblock Complex shapes self-assembled from single-stranded {DNA} tiles.
\newblock {\em Nature}, 485(7400):623--626, 2012.

\bibitem{winfree1998design}
Erik Winfree, Furong Liu, Lisa~A Wenzler, and Nadrian~C Seeman.
\newblock Design and self-assembly of two-dimensional {DNA} crystals.
\newblock {\em Nature}, 394(6693):539--544, 1998.

\bibitem{woo2011programmable}
Sungwook Woo and Paul~WK Rothemund.
\newblock Programmable molecular recognition based on the geometry of {DNA}
  nanostructures.
\newblock {\em Nature chemistry}, 3(8):620, 2011.

\bibitem{drmaurdsa}
Damien Woods, David Doty, Cameron Myhrvold, Joy Hui, Felix Zhou, Peng Yin, and
  Erik Winfree.
\newblock Diverse and robust molecular algorithms using reprogrammable {DNA}
  self-assembly.
\newblock {\em Nature}, 567:366--372, 2019.
\newblock \href {https://doi.org/10.1038/s41586-019-1014-9}
  {\path{doi:10.1038/s41586-019-1014-9}}.

\bibitem{nupack}
Joseph~N. Zadeh, Conrad~D. Steenberg, Justin~S. Bois, Brian~R. Wolfe,
  Marshall~B. Pierce, Asif~R. Khan, Robert~M. Dirks, and Niles~A. Pierce.
\newblock Nupack: Analysis and design of nucleic acid systems.
\newblock {\em Journal of Computational Chemistry}, 32(1):170--173, 2011.
\newblock URL: \url{https://onlinelibrary.wiley.com/doi/abs/10.1002/jcc.21596},
  \href
  {http://arxiv.org/abs/https://onlinelibrary.wiley.com/doi/pdf/10.1002/jcc.21596}
  {\path{arXiv:https://onlinelibrary.wiley.com/doi/pdf/10.1002/jcc.21596}},
  \href {https://doi.org/10.1002/jcc.21596} {\path{doi:10.1002/jcc.21596}}.

\bibitem{zhang2015complex}
Fei Zhang, Shuoxing Jiang, Siyu Wu, Yulin Li, Chengde Mao, Yan Liu, and Hao
  Yan.
\newblock Complex wireframe {DNA} origami nanostructures with multi-arm
  junction vertices.
\newblock {\em Nature nanotechnology}, 10(9):779, 2015.

\end{thebibliography}

\end{document}